\documentclass[a4paper, 10pt]{article}
\usepackage[utf8]{inputenc}
\usepackage[T1]{fontenc}
\usepackage{todonotes}
\usepackage{subcaption}
\usepackage{authblk}
\usepackage{amsmath,amssymb}

\title{Unique in the Smart Grid - The Privacy Cost of Fine-Grained Electrical Consumption Data}

\author[1,3,*]{
    Antonin Voyez
}
\author[1]{Tristan Allard}
\author[1,2]{Gildas Avoine}
\author[3]{Pierre Cauchois}
\author[1]{Elisa Fromont}
\author[4]{Matthieu Simonin}
\affil[1]{Univ Rennes, CNRS, IRISA, France}
\affil[2]{INSA Rennes, CNRS, IRISA, France}
\affil[3]{Enedis, France}
\affil[4]{Inria, IRISA, France}

\affil[*]{All requests should be addressed to A. V. (antonin.voyez@irisa.fr)}

\begin{document}

\maketitle

\thispagestyle{empty}

\section*{Introduction}

According to the International Energy Agency (IEA), \emph{smart grids} are essential for coping with the hard challenges of the 21st century related to energy\cite{oecd_2011_smart_meter} (e.g., energy security, economic development, climate change mitigation).
Within a smart grid, \emph{smart meters} are key devices installed on households (or companies) side. They are able to measure at a high rate the electric energy consumed (e.g., the smart meters used in this study currently perform up to one measure every $30$ minutes) and to report the resulting sequences of measures, called \emph{electric consumption time series} below, to the grid manager. 
During the last decade, many grid managers have deployed nationwide networks of smart meters.
For example, in the European Union, more than 200 million smart meters~\cite{jrc_2022_smartmeter} have been deployed including more than 90 \% of the French households (i.e., 30 million households)~\cite{cre_2020_activity}, and more than 100 million meters have been deployed in the United States~\cite{eia_2020_smart_meter}.

Fine-grained electric consumption time series yield high values both for the internal grid actors (i.e., the grid manager, the electricity producers, the electricity consumers) but also for external actors~\cite{cre_smart_grids, oecd_2018_smart_grid}.
Strongly encouraged by modern laws related to open data (for example, the European open data directive identifies energy consumption data as a high-value dataset for society\cite{regulation_2019-euopendata}), grid operators are launching ambitious data sharing programs for making electric consumption time series available to the general public, either to selected participants through \emph{closed data contracts}~\cite{pge_opendata} or following \emph{open data principles}~\cite{enedis_opendata}.% 

However, the wealth of information carried by fine-grained electric consumption time series is a double-edged sword: it is a direct threat to individuals right to privacy. %
Non-Intrusive Load Monitoring techniques (NILM)~\cite{hart1992nonintrusive, DBLP:conf/gi/KlemenjakG16} aim at extracting valuable information about households from fine-grained electric time series. They have provided extensive arguments supporting this claim along the past decades by, e.g., identifying the devices used by the household~\cite{birt2012disaggregating, rottondi2019optimisation}, computing the household occupancy\cite{chen2013non, becker2018exploring}, computing socio-economical metrics related to the households~\cite{beckel2013automatic} and even the unemployment status of the related individuals~\cite{montanez2020machine}. %
There is no doubt that the current NILM studies, with limited access to electrical consumption time series and to ground truth information about individuals, only scratch the surface of the information that can be inferred from electric consumption time series. %
Electric consumption time series obviously fall within the scope of modern personal data protection laws such as GDPR\cite{regulation_2016_gdpr} or CCPA\cite{regulation_2018_ccpa} which impose grid operators to provide robust privacy guarantees while sharing the electric consumption time series collected. %
Would that be enough then to naively remove personally identifiable information from the electrical consumption time series (e.g., the meter identifier, the name of the corresponding individuals) for publishing them without jeopardizing privacy? (Similarly to a pseudonymization scheme.) In other words, would the link between individuals and their time series (and consequently with the various information that might be inferred from the time series) be deleted? In this paper, we provide for the first time a detailed answer to this question. %
While electric consumption datasets~\cite{noauthor_issda_nodate, ukpn_london_2013} are already published using simple pseudonymisation scheme, the answer is no: mass re-identification is possible even when the electric consumption time series are dramatically degraded or aggregated along the time. 

\emph{Uniqueness} is a widely used measure for evaluating the vulnerability to re-identifications of personal data.
For example, the famous re-identification of Governor Weld performed by Sweeney within a health dataset two decades ago~\cite{sweeney2002k} was allowed by the fact that Governor Weld's record was unique on the $\mathtt{(date~of~birth,~zip code,~sex)}$ columns available in the health dataset disclosed: any adversary knowing the $\mathtt{(date~of~birth,~zipcode,~sex)}$ triple of Governor Weld was able to join it with the dataset in order to obtain his health information. Obviously, the higher the rate of unique individuals' records is, the more vulnerable to re-identifications the dataset is. Uniqueness was used formally in the early Netflix attack~\cite{narayanan2008robust}, exploited then by large scale empirical studies on mobility traces~\cite{de2013unique} and transaction data~\cite{de2015unique}, and is now well recognised as a risk measurement method~\cite{sekara2021temporal, romanini2021privacy}. %
Beyond re-identification, many attacks\cite{jawurek2011smart, buchmann2013re, dietrich2020lack, DBLP:conf/ndss/PyrgelisTC18, buescher2017two, watson2022on}, such as the membership inferences ones, are expected to perform better on outliers (i.e., unique?) data. 
 
In this paper, we study the uniqueness of fine-grained electric consumption time series at a nationwide scale in order to evaluate the vulnerability to re-identifications of large scale datasets collected through smart meters.
Uniqueness quantifies, in this context, the fraction of households that can be re-identified by accessing the electric consumption time series (see the Methods Section for details on the uniqueness computation). %
%%%% Début ajout paragraphe DP
Note that using strong privacy-preserving data publishing methods such as $\epsilon$-differential privacy~\cite{dwork2006calibrating} would straightforwardly prevent re-identification attacks.
Indeed, $\epsilon$-differentially private algorithms would not disclose raw time series, but would rather output aggregated and perturbed information\cite{rastogidp-sigmod10} that are immune to re-identifications. %
However, $\epsilon$-differentially private algorithms calibrate the magnitude of the perturbation according to the quantity of information related to each single individual (e.g., the length of electric consumption time series).
In a context where the time series are unbounded, maintaining high utility without compromising the privacy level seems to be paradoxical. %

This current lack of ideal privacy protection measure for highly valued time series data is %another 
both a strong argument towards the systematic study of privacy threats caused by high data uniqueness and a call towards the development of privacy-preserving data publishing methods dedicated to unbounded time series. %

This work is, to the best of our knowledge, the first uniqueness study performed at a real-life nationwide scale and dedicated to fine-grained electric consumption time series. %
Previous large scale uniqueness studies~\cite{narayanan2008robust, de2013unique, de2015unique} do not include electric consumption time series, while related works focusing on the vulnerabilities to re-identification of electric consumption time series~\cite{jawurek2011smart, buchmann2013re, dietrich2020lack, buescher2017two} do not perform any in-depth uniqueness study.
A recent work~\cite{Creu2022InteractionDA} focuses on re-identifying individuals when the adversarial background knowledge is at a time period different from the time period of the dataset disclosed, but uniqueness is not studied and the dataset consists in interaction data between individuals (e.g., who calls who and when). %

Our results are based on two real-life large scale electric consumption time series datasets collected over a year by our industrial partner Enedis, the French national electricity distribution system operator. %
In a nutshell, we observe that even at such a large-scale, almost all electric consumption time series are unique with respect to a handful of consecutive electric consumption measures.
For example, on average, 90\% of the time series can be uniquely re-identified by an adversary knowing only 5 consecutive daily measures.  
Moreover, tremendously degrading the precision of the electric consumption measures is not enough to hinder possible re-identifications, even when rounding the measures from the watt to the kilowatt on half-hourly consumption measures. %
Note that the fewer the electric consumption measures needed to be unique, the easier it is for an attacker to perform a re-identification. %

\section*{Methods}

\subsection*{Datasets}
% 175  words
This work was conducted with our industrial partner, the French national electricity distribution system operator, called Enedis, using two nationwide electric consumption datasets. These datasets leverage the recent French nationwide deployment of electric smart meters. An electric smart meter is a physical device installed at the entry point of the household electrical network that measures the electric power at a given rate (i.e., currently every $30$ minutes in France) and at a given precision (e.g., to the watt in France, and up to $36,000$~W), and that sends it to a central server.
The two datasets contain electric consumption time series collected  between September 2020 and September 2021 from residential smart meters at two distinct time scales.
The first dataset, called \emph{the daily dataset}, contains the daily electric consumption measures (in watt-hour -- Wh) of around 25M residential meters.
The second dataset, called \emph{the half-hourly dataset}, contains the half-hourly electric power measures (in watt -- W) of around 2.5M residential meters. %
In the other sections, we use the term electric consumption measures for both datasets. %
The data used in this analysis has been collected and processed by fully complying with the current legislation in France and in the European Union~\cite{regulation_2016_gdpr, regulation_2021_frenergy} (i.e., the GDPR).
In particular, Enedis collected the consent of the individuals concerned as part of its regular collection process. The data is stored and analyzed on the Enedis computing infrastructure. %
Due to the sensitivity of electric consumption time series, we are prohibited by laws from making our datasets public. % 
Any data access request must be addressed to the Enedis company. %

\subsection*{Uniqueness and Entropy}

We apply the same method for computing the uniqueness regardless of the dataset. %(e.g., daily dataset, half-hourly dataset).
Given $k$, we compute the uniqueness at each timestamp $t$ by considering for each time series the sub-sequence starting at $t$ included and containing $k$ measures (also called a \emph{$k$-length sub-sequence} below) and by computing the uniqueness in the resulting set of sub-sequences. Finally, for a given $k$, we compute and plot the average and min / max interval uniqueness over all timestamps. %

Let $S$ be a set of $n$ electrical consumption time series with $m$ timestamps each (e.g., the daily dataset). Given $k$, we denote $S_t^k$ the dataset derived from $S$ that contains for each time series the $k$ consecutive timestamps starting at time $t$ included (and consequently ending at time $t+k-1$). Note that we ignore the $k$-length sub-sequences that have missing values (due to, e.g., transmission errors). % 
Let $\mathtt{U}$ be the function that, given as input a dataset $S_t^k$, outputs the set of time series that are unique in $S_t^k$: $s \in \mathtt{U}(S_t^k) \iff \nexists s' \in S_t^k \sim \{s\}$ %
s.t. $s' = s$. %
Finally, given a derived dataset $S_t^k$, the uniqueness at time $t$ for $k$-length sub-sequences, denoted $u_t^k$, is the fraction of unique time series in $S_t^k$: $u_t^k = |\mathtt{U}(S_t^k)| / |S_t^k|$. 

The entropy is computed using the Shannon entropy~\cite{shannon1948mathematical} formulae applied to $k$-length sub-sequences.
At each timestamp $t$ the entropy is defined as follow: $e_t = -\sum_{i} P_i \log_2 P_i$, with $P_i$ defined as the number of times the $i^{th}$ $k$-length sub-sequence appears in $S^k_t$ divided by the total number of $k$-length sub-sequences in $S^k_t$. % 

Uniqueness and entropy are computed at every timestamp of the year and for various values of $k$. %

\subsection*{Data Degradation}

We also study the impact on uniqueness of a more and more severe information loss. We degrade our time series by rounding their values by $1$, $2$, and $3$ orders of magnitude (i.e., respectively to the closest 10 W (or Wh), the closest 100 W (or Wh), and the closest kW (or kWh)) and compute uniqueness as explained above. To this end, consumption measures are rounded before computing uniqueness or entropy through the standard SQL \texttt{ROUND} function.% is used.

\section*{Results}

\begin{figure}[h]
\centering

    \begin{subfigure}[b]{0.33\textwidth}
        \centering
        \includegraphics[width=1\linewidth]{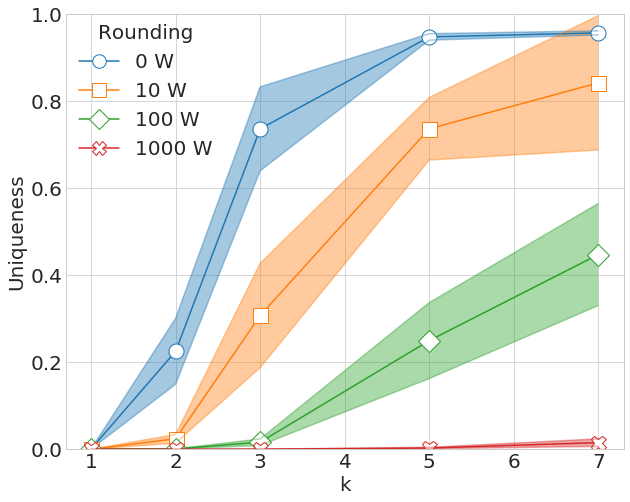}
        \caption{}
        \label{fig:uni_rate_hh}
    \end{subfigure}\hfill%
    \begin{subfigure}[b]{0.33\textwidth}
        \centering
        \includegraphics[width=1\linewidth]{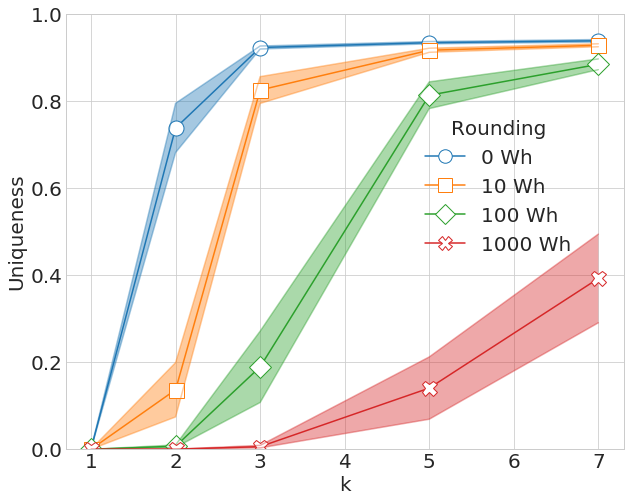}
        \caption{}
        \label{fig:uni_rate_daily}
    \end{subfigure}\hfill%
    \begin{subfigure}[b]{0.33\textwidth}
        \centering
        \includegraphics[width=1\linewidth]{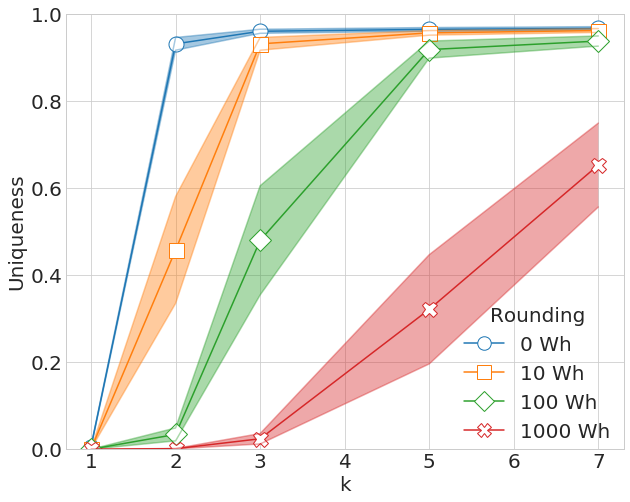}
        \caption{}
        \label{fig:uni_rate_daily_cdc}
    \end{subfigure}
    
    \caption{
        Average uniqueness (mean with minimum / maximum uniqueness observed) according to $k$ and to the rounding for (a) the half-hourly dataset, (b) the daily dataset and (c) the subset of the daily dataset restricted to time series generated by the smart meters having generated the half-hourly dataset. %
    }
    \label{fig:uni_rate}
\end{figure}

We show in Figure~\ref{fig:uni_rate} how uniqueness varies, for our two datasets, according to $k$ and to the order of magnitude of the rounding.
A very high uniqueness (above 90 \%) is reached at $k=5$ (i.e., 2h30 consumption in the half-hourly dataset (Figure~\ref{fig:uni_rate_hh}), and 3 days consumption in the daily dataset (Figure~\ref{fig:uni_rate_daily} and \ref{fig:uni_rate_daily_cdc})), without any rounding. 
It is worth noting that sub-sequences of length $k=1$ have a very low uniqueness: almost every single smart meter consumption is shared by at least another meter.
However, as time goes by, even by a few hours, almost every smart meter generates a different consumption time series, making them unique. 

\begin{figure}[h!]
    \centering
    
    \begin{subfigure}[b]{0.48\textwidth}
        \centering
        \includegraphics[width=0.95\linewidth]{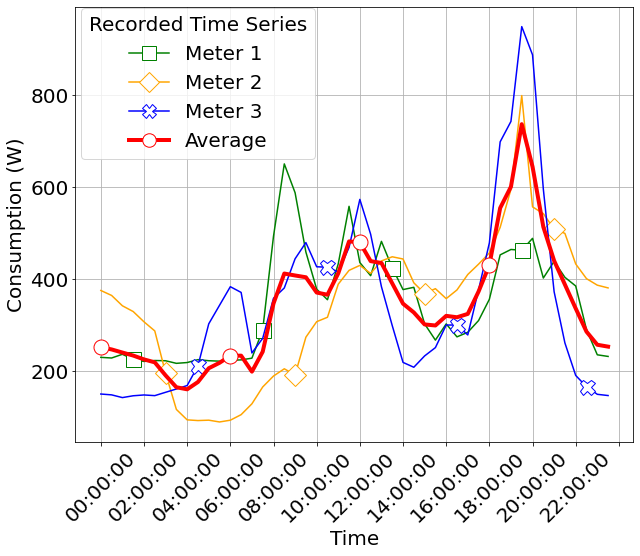}
        \caption{}
        \label{fig:exemple_cdc}
    \end{subfigure}\hfill%
    \begin{subfigure}[b]{0.48\textwidth}
        \centering
        \includegraphics[width=1\linewidth]{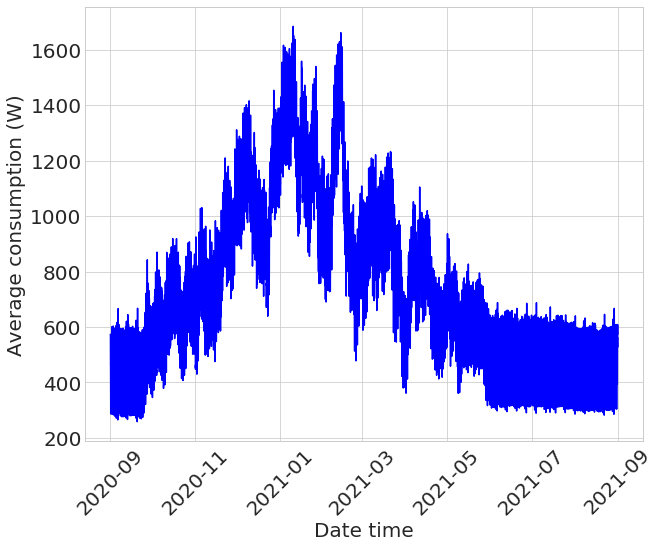}
        \caption{}
        \label{fig:cdc_avg}
    \end{subfigure}

    \caption{
    (a) Three illustrative electric consumption time series (during a single day at a $30$ minutes rate) and the time series containing for each timestamp $t$ the average of the three series at $t$.
    (b) Average of the time series of the half-hourly dataset (for each timestamp $t$, average of the series at $t$).
    %\comment{Moving theses figures (previously Figure 1)}
    }
    \label{fig:exemple_cdcs}
\end{figure}

Focusing on Figure~\ref{fig:uni_rate_hh}, i.e., the uniqueness of the half-hourly dataset, we confirm the intuition that the higher the rounding the lower the uniqueness, which nevertheless remains far from being negligible even with the strongest degradation. For example, rounding to $2$ orders of magnitude (i.e., to the closest 100 W) results in 40\%+ uniqueness for $k = 7$, and rounding to $3$ orders of magnitude (i.e., to the closest kW -- a very strong degradation given the average consumption and the standard deviation -- see Figure~\ref{fig:cdc_avg}), the uniqueness rises to almost $0.5$\% which corresponds to 12,500 unique time series: a non-negligible number of households at risk. Increasing $k$ might obviously increase uniqueness. 
Note that such a rounding further results in a dramatic information loss. Indeed, although the definition domain of our electric consumption measures is $[0;~36,000]~W$, the vast majority of measures fall between $500~W$ and $1,500~W$, as illustrated by Figure~\ref{fig:cdc_avg}).
Overall, the average consumption measure over the full year is 725~W with a mean standard deviation of 950~W. 

Focusing on Figure~\ref{fig:uni_rate_daily} now, i.e., the uniqueness of the daily dataset, we observe even higher uniqueness, whether rounding is enabled or not. Because rounding applies here to much higher measures in expectation (daily measures instead of half-hourly measures), it has (unsurprisingly) much less impact on uniqueness: after rounding to the kW the daily measures, around 40~\% of the 25M meters are still unique with $k=7$ (i.e., one week rounded daily consumption). %
Figure~\ref{fig:uni_rate_daily_cdc} shows the uniqueness of a subset of the daily dataset including only the daily time series of the 2.5M smart meters involved in the half-hourly dataset. By comparing it to Figure~\ref{fig:uni_rate_daily} we observe the impact on the uniqueness of the scale of the dataset. In general, thanks to its sparser space, the subset of the daily dataset reaches higher uniqueness. 
More precisely, increasing the number of time series by an order of magnitude -- i.e., from 2.5M to 25M -- slightly reduces uniqueness: obtaining 90~\% uniqueness requires $k=3$ consecutive measures without rounding (instead of $k=2$), requires $k=5$ consecutive measures with a rounding to the 10~Wh (instead of $k=3$), and requires $k=7$ with a rounding to the 100~Wh (instead of $k=5$). With the highest rounding enabled (to the kWh) and when $k=7$, uniqueness drops from around 70~\% on the daily subset to 40~\% on the full daily dataset.

Note that the min and max uniqueness may vary more or less depending on $k$ and on the order of magnitude of rounding. For example, for the half-hourly dataset, with $k=3$ and no rounding,
the min and max uniqueness differ by $40$~\% at most while for most parameter settings, the difference between the min and max uniqueness remain small.

Finally, we observe that uniqueness reaches a maximum value (e.g., around 96~\% for both the half-hourly dataset and the daily dataset). Once this value is reached, increasing $k$ only increases the uniqueness by a tiny fraction. This can be explained by the significant proportion of electric consumption measures equal to $0~W$ (between 3\% and 10\%).

To illustrate why uniqueness reaches such high values in our datasets we consider the three illustrative electric consumption time series shown in  Figure~\ref{fig:exemple_cdc}. Each series corresponds to a distinct smart meter (i.e., a distinct household) measuring power consumption in W., for a single day with 48 measures (i.e., $30$ minutes rate).
The three series \emph{look} similar because many individuals share common activities (e.g., waking up, commuting to work, performing other activities roughly at the same time) and their electric consumption time series mirror these common behaviors.
However, no household is the exact copy of another.
First the same activity of an electric device will be most likely shifted along the time axis due to difference in household schedules.
Also using distinct electrical devices results in distinctive electric consumption.
This contributes to making them \emph{unique along the time}.

\begin{figure}[h]
\centering

    \begin{subfigure}[b]{0.48\textwidth}
        \centering
        \includegraphics[width=1\linewidth]{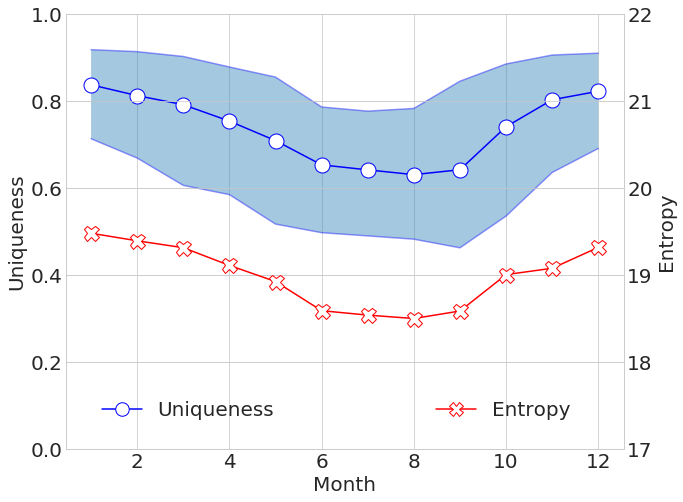}
        \caption{}
        %\caption{Uniqueness and entropy per month.}
        \label{fig:pattern_montly}
    \end{subfigure}\hfill%
    \begin{subfigure}[b]{0.48\textwidth}
        \centering
        \includegraphics[width=1\linewidth]{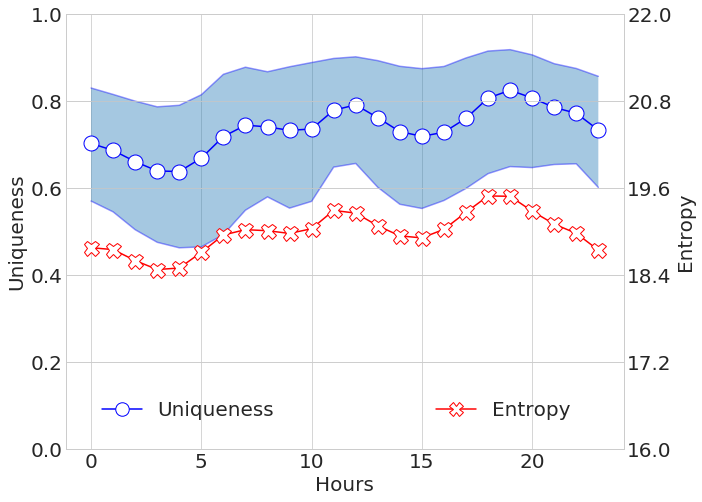}
        \caption{}
        %\caption{Uniqueness and entropy per hours.}
        \label{fig:pattern_hourly}
    \end{subfigure}\hfill%
    \begin{subfigure}[b]{0.3\textwidth}
        \centering
        \includegraphics[width=1\linewidth]{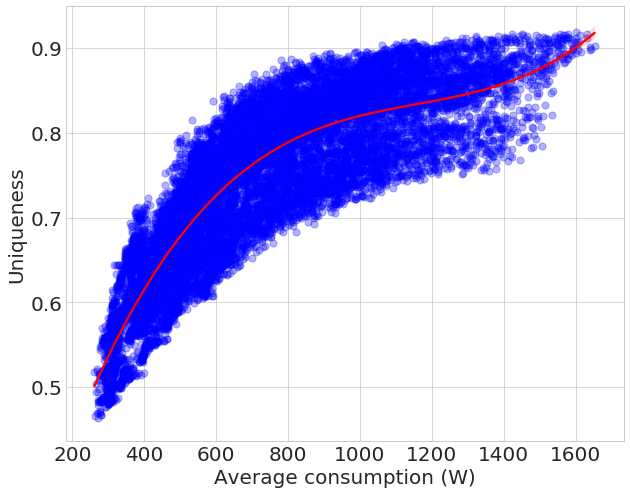}
        \caption{}
        %\caption{Uniqueness function of the consumption}
        \label{fig:uni_conso}
    \end{subfigure}\hfill%
    \begin{subfigure}[b]{0.3\textwidth}
        \centering
        \includegraphics[width=1\linewidth]{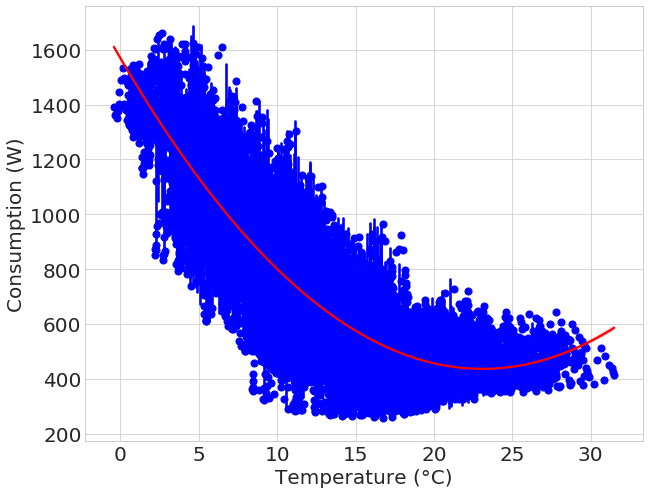}
        \caption{}
        %\caption{Consumption function of the temperature}
        \label{fig:conso_temp}
    \end{subfigure}\hfill%
    \begin{subfigure}[b]{0.3\textwidth}
        \centering
        \includegraphics[width=1\linewidth]{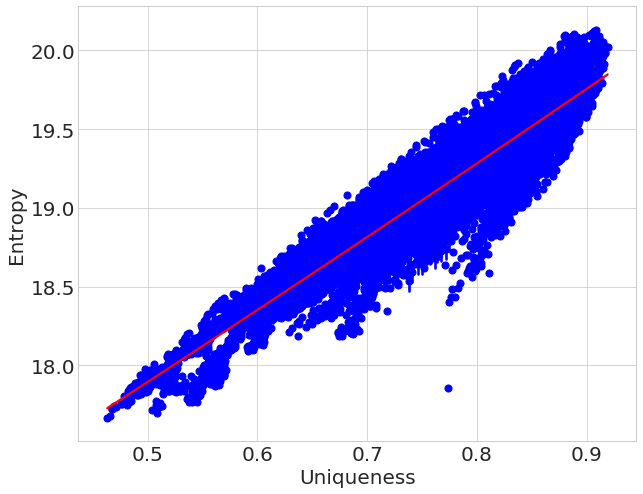}
        \caption{}
        %\caption{Uniqueness function of the entropy.}
        \label{fig:conso_entro}
    \end{subfigure}
    
    \caption{
        Monthly (a) and hourly (b) uniqueness along time (mean with the minimum / maximum uniqueness observed)}, together with the relationship between electric consumption and uniqueness (c), between temperatures and electric consumption (d), and between entropy and uniqueness (e). All graphs are computed based on the half-hourly dataset with $k=3$ and no rounding (i.e., to the W). 
    \label{fig:pattern}
\end{figure}

Figure~\ref{fig:pattern} shows how uniqueness varies with time.
In particular, Figure~\ref{fig:pattern_montly} shows a first seasonal variation leading to higher uniqueness in winter than in summer.
Figure~\ref{fig:pattern_hourly} shows that uniqueness is higher during the day than during the night with two peaks, the first peak at noon (12h) and the second peak in the evening (18h-22h -- i.e., 6PM-10PM).
The entropy of the sets of $k$-length subsequences at each timestamp are also plotted on Figures~\ref{fig:pattern_montly} and~\ref{fig:pattern_hourly} in order to observe visually the correlation between uniqueness and entropy, also clearly illustrated in Figure~\ref{fig:conso_entro}. The Pearson correlation coefficient between uniqueness and entropy is $0.94$ (the closer to $1$ the stronger the correlation).
In addition, as shown in Figures~\ref{fig:uni_conso} and~\ref{fig:conso_temp}, uniqueness is also strongly correlated to the average consumption (Pearson coefficient of $0.8$) which is itself strongly (negatively) correlated to the temperatures (Pearson coefficient of $-0.8$): the lower the temperature the higher the consumption, and the higher the consumption the higher the uniqueness. %
The consumption seems to stabilize for an average temperature above 15 °C. %
At local levels, a temperature above 30 °C leads to a consumption bounce. However, France is not yet warm enough to be able to spot such patterns at a nationwide scale.
In a nutshell, uniqueness, and consequently re-identification risks, are higher for a winter evening than for a summer night. %

\section*{Discussion}

In this study, we present the first uniqueness results computed over two large-scale, nationwide, electric consumption time series datasets. We show that uniqueness reaches dramatically high values even when computed over very small sub-sequences.
In other words, our results show that re-identifying millions of households is possible in large-scale electric consumption time series datasets given a tiny sub-sequence of the full time series.
Our results hold despite the following adverse facts.
First, individuals tend to behave similarly (e.g., sleeping patterns, commuting times). While this (relative) uniformity can be observed in the results that only include single consumption measures ($k=1$), uniqueness increases fastly when considering only a little bit more consumption measures ($k>1$).
Second, individual measures are concentrated over a small part of the definition domain. Indeed, Figure~\ref{fig:cdc_avg} illustrates the fact that, in general, half-hourly measures fall between $500~W$ and $1,500~W$ whereas the definition domain extends to $36,000~W$. Again, this impacts our results when uniqueness is computed on single electric consumption measures, but considering only a few additional measures make uniqueness rise tremendously.
Third, we compute uniqueness at each timestamp $t$ based on $k$-length sub-sequences, i.e., sub-sequences of $k$ consecutive consumption measures starting at time $t$. Performing an exhaustive search of the subsets of $k$ measures, not necessarily consecutive, that lead to the highest uniqueness might indeed lead to higher uniqueness. However, we believe that the uniqueness results that we obtain are already sufficiently high for raising strong concerns about the re-identifiability of households within electric consumption time series datasets.
Fourth, to the best of our knowledge, our datasets are the largest electric consumption time series datasets considered to date (i.e., around 2.5M half-hourly time series and around 25M daily time series -- both during one year). Despite the possible collisions, that increase with the number of time series in the dataset, uniqueness remains dramatically high, even when considering small sub-sequences. 

Our study also shows the impact on uniqueness of degrading severely the time series. Surprisingly, uniqueness remains high even when losing orders or magnitude precision and when considering small sub-sequences.
Even when uniqueness drops -- i.e., when rounding the half-hourly series to the kW ($3$ orders of magnitude rounding, resulting in information loss hard to cope with for many applications) -- around $12,500$ thousands of households among the 2.5M of the full half-hourly dataset remain at risk of being re-identified. %
This shows the limitation of naive protection methods (e.g., rounding) from re-identification attacks. 

The high uniqueness depicted in the results of this study shows that re-identification of households in large-scale electric consumption datasets is possible with high probability by adversaries knowing only a small subset of consumption measures of their target(s). Detailed electrical consumption measures can be captured today by a wide range of actors in addition to the grid manager. First, electricity providers obviously need to charge their customers but the information they access is often more detailed than necessary (e.g., daily consumption measures, half-hourly measures). This depends, e.g., on legal restrictions, or on the individuals consent if the legal framework requires it (e.g., GDPR). Second, smart plugs (or similar devices) allow individuals to monitor over the Internet the power consumption of elected devices. Electric consumption measures are thus sent to third-party information systems and monitored by the individuals through smartphone applications. Third, the grid manager may also allow individuals to monitor their electric consumption time series through a user account on a dedicated web portal. Although all these actors may not adopt adversarial behaviors (especially the individuals monitoring their own consumption measures), the attack surface is large, with weaknesses, and they may suffer from negligence. This increases the risk of leaking electric consumption measures to the wild.
Additionally, as shown by our uniqueness results on severely degraded electric consumption time series, even rough estimations of electric consumption measures might be sufficient for performing a re-identification (e.g., based on other sources of information about the household, based on past data, based on a subset of the electric consumption): adversaries with approximate knowledge might still be able to benefit from the high uniqueness of degraded electric consumption time series for performing re-identifications. 

%% Capacité à généraliser les résultats
Our uniqueness results are based on two French nationwide electric consumption datasets.
While we are aware that biases might impact the actual uniqueness numbers (e.g., climate biases, socio-cultural biases, political biases), we believe that similar conclusions about uniqueness can be drawn from a large number of other nationwide electric consumption datasets.
Indeed, climate is the main driving force for households energy consumption and consequently for uniqueness (temperatures in particular -- see the Results Section for details).
However, our results show that for $k \geq 5$, worryingly high uniqueness levels are reached (above 90\%), independently from local variations of entropy and of climate.

We further note that although the trends are similar, higher uniqueness can be observed on the (small scale) publicly available CER ISSDA electric consumption time series dataset~\cite{noauthor_issda_nodate}.
This can be explained by the small size of the dataset which results in a much sparser dataset, thus in less collisions, and as a result in higher uniqueness. % 

As the importance of the collection of fine grained electrical consumption data grows, our study highlights the privacy vulnerability of publishing sets of individual consumption series.
We show that almost every participant are unique by knowing a handful of points, even an important reduction in the records precision is not enough to significantly reduce uniqueness while keeping the data usability.
Uniqueness is function of the dataset entropy and is influenced by external factors such as the temperature and is strongly correlated with the dataset entropy.
Overall, our uniqueness study performed over large, real-life, electricity consumption dataset show high unicity and therefore potential privacy threats.

\bibliographystyle{unsrt}
\bibliography{biblio}

\section*{Acknowledgements}
We would like to thank the members of the TAILOR (H2020 ICT-48) network for some very fruitful discussions about privacy risks.

\section*{Author contributions statement}

A.V., T.A., G.A., P.C., E.F., M.S. designed the study, analysed the results and wrote the paper.
A.V., P.C., M.S. designed and wrote the experiment code and analysis tools.

\section*{Additional information}

\textbf{Competing interest} 
This research has been partially funded by Enedis. A.V. and P.C. are Enedis employees. The rest of the authors have no any competing interest.\\
\textbf{Data availability}  Due to the sensitivity of electric consumption time series, we are prohibited by laws from making our datasets public. % 
Any data access request must be addressed to the Enedis company.\\
\textbf{Correspondence} and requests should be addressed to A.V.

\end{document}